\documentclass[aps,prl,twocolumn,showpacs,superscriptaddress,floatfix]{revtex4-1}

\usepackage{graphicx}
\usepackage{dcolumn}
\usepackage{bm}
\usepackage{amssymb}
\usepackage{microtype}
 
\newcommand{\pyro}[2]{#1$_2$#2$_2$O$_7$}
\newcommand{\ket}[1]{\ensuremath{|#1\rangle}}

\begin{document}

\title{Magnetic Frustration in Lead Pyrochlores}

\author{A.~M.~Hallas}
\affiliation{Department of Physics and Astronomy, McMaster University, Hamilton, ON, L8S 4M1, Canada}

\author{A.~M.~Arevalo-Lopez}
\affiliation{Centre for Science at Extreme Conditions and School of Chemistry, University of Edinburgh, Peter Guthrie Tait Road, King's Buildings, Edinburgh EH9 3FD, United Kingdom}

\author{A.~Z.~Sharma}
\affiliation{Department of Chemistry, University of Manitoba, Winnipeg, MB, R3T 2N2, Canada}

\author{T.~Munsie}
\affiliation{Department of Physics and Astronomy, McMaster University, Hamilton, ON, L8S 4M1, Canada}

\author{J.~P.~Attfield}
\affiliation{Centre for Science at Extreme Conditions and School of Chemistry, University of Edinburgh, Peter Guthrie Tait Road, King's Buildings, Edinburgh EH9 3FD, United Kingdom}

\author{C.~R.~Wiebe}
\affiliation{Department of Physics and Astronomy, McMaster University, Hamilton, ON, L8S 4M1, Canada}
\affiliation{Department of Chemistry, University of Manitoba, Winnipeg, MB, R3T 2N2, Canada}
\affiliation{Department of Chemistry, University of Winnipeg, Winnipeg, MB, R3B 2E9 Canada}
\affiliation{Canadian Institute for Advanced Research, 180 Dundas St. W., Toronto, ON, M5G 1Z7, Canada}

\author{G.~M.~Luke}
\affiliation{Department of Physics and Astronomy, McMaster University, Hamilton, ON, L8S 4M1, Canada}
\affiliation{Canadian Institute for Advanced Research, 180 Dundas St. W., Toronto, ON, M5G 1Z7, Canada}

\date{\today}

\begin{abstract}
The rich phase diagrams of magnetically frustrated pyrochlores have maintained a high level of interest over the past 20 years. To experimentally explore these phase diagrams requires a means of tuning the relevant interactions. One approach to achieve this is chemical pressure, that is, varying the size of the non-magnetic cation. Here, we report on a new family of lead-based pyrochlores \pyro{A}{Pb} (A = Pr, Nd, Gd), which we have characterized with magnetic susceptibility and specific heat. Lead is the largest known possible B-site cation for the pyrochlore lattice. Thus, these materials significantly expand the phase space of the frustrated pyrochlores. \pyro{Pr}{Pb} has an absence of long-range magnetic order down to 400~mK and a spin ice-like heat capacity anomaly at 1.2~K. Thus, \pyro{Pr}{Pb} is a candidate for a quantum spin ice state, despite weaker exchange. \pyro{Nd}{Pb} transitions to a magnetically ordered state at 0.41~K. The Weiss temperature for \pyro{Nd}{Pb} is $\theta_{\text{CW}}$~=~$-$0.06~K, indicating close competition between ferromagnetic and antiferromagnetic interactions. \pyro{Gd}{Pb} is a Heisenberg antiferromagnet that transitions to long-range magnetic order at 0.81~K, in spite of significant site mixing. Below its ordering transition, we find a $T^{3/2}$ heat capacity dependence in \pyro{Gd}{Pb}, indication of a ground state that is distinct from other gadolinium pyrochlores. These lead-based pyrochlores provide insight into the effects of weakened exchange on highly frustrated lattices and represent further realizations of several exotic magnetic ground states which can test theoretical models. 
\end{abstract}

%\pacs{75.30.Cr, 75.40.Cx, 75.50.Lk}% PACS, the Physics and Astronomy

\maketitle

\section{Introduction}
Magnetic frustration occurs when it is impossible to satisfy all pairwise magnetic interactions simultaneously and is typically a consequence of the underlying lattice geometry. The canonical example of magnetic frustration is antiferromagnetic Ising spins placed on the vertices of an equilateral triangle; in this scenario, only two of the three spin pairs can be anti-parallel, resulting in magnetic frustration. Most structures studied in the context of magnetic frustration are composed of triangular motifs \cite{B003682J}. Some examples of frustrated lattice geometries include the kagom\'{e}, honeycomb, and perovskite lattices. However, there is perhaps no lattice better suited to the study of magnetic frustration than the pyrochlore lattice \cite{GGG}. Pyrochlores, with formula \pyro{A}{B}, are composed of two interpenetrating networks of corner-sharing tetrahedra. The vertices of one tetrahedral network are occupied by the A$^{3+}$ cations while the second network is occupied by the B$^{4+}$ cations. This corner-sharing tetrahedral arrangement is subject to extreme frustration when occupied by a magnetic ion. It is fortunate for the study of magnetic frustration that a large fraction of the cations which can occupy the pyrochlore lattice are, in fact, magnetic.

The most well-characterized family of frustrated pyrochlores are the rare earth titanate pyrochlores, \pyro{R}{Ti} \cite{GGG}. The B-site cation, Ti$^{4+}$, has a closed valence shell and is thus non-magnetic. By varying the A-site rare earth cation, many unique and exotic magnetic ground states have been uncovered in this family. The spin ice state, a magnetic analog for the proton arrangement in water ice, is observed for R~=~Ho \cite{Harris1997}, Dy \cite{Ramirez1999}. WWhen the rare earth site is occupied by \emph{R} = Tb, there is an absence of conventional long-range magnetic order down to 70~mK, leading Tb$_2$Ti$_2$O$_7$ to be considered a spin liquid candidate \cite{PhysRevB.68.180401,PhysRevLett.82.1012}. Long range order may also be absent in the case of the apparent quantum spin liquid R~=~Yb \cite{PhysRevLett.88.077204,PhysRevB.70.180404}. A detailed investigation of the Hamiltonian of \pyro{Yb}{Ti} has justified the label ``quantum spin ice'' \cite{PhysRevX.1.021002}. However, the low temperature magnetism of \pyro{Yb}{Ti} remains controversial, due to sample variation \cite{PhysRevB.84.172408,PhysRevB.86.174424} and conflicting reports \cite{YbTiONatComm,PhysRevB.88.134428}. Even the titanate pyrochlores that transition to long-range magnetic order, R~=~Gd, Er, have exotic properties. In the case of Er$_2$Ti$_2$O$_7$, one proposed scenario is that the transition to long-range magnetic order is selected by thermal fluctuations (i.e., order-by-disorder) \cite{PhysRevB.68.020401,PhysRevLett.109.167201,PhysRevB.88.220404}. In \pyro{Gd}{Ti}, an uncommon example of Heisenberg spins on a pyrochlore lattice \cite{PhysRevB.59.14489}, the antiferromagnetic ground state is a complex multi-$k$ structure with a two-stage ordering \cite{PhysRevB.64.140407,0953-8984-16-28-L01}. 

The plethora of exotic magnetism discovered in the titanate pyrochlores was further expanded through substitution of titanium for other non-magnetic cations of varying size, \emph{i.e.}, chemical pressure \cite{Stannates,doi:10.1021/ic50066a038}. Studies with chemical pressure have allowed investigation into the phase space for these frustrated materials. In the pyrochlores, variation of the non-magnetic B-site cation has two main consequences. Firstly, the change in ionic radii introduces a positive or negative chemical pressure. This results in differences in the distances between magnetic A-site cations, allowing an effective tuning of the exchange interaction \cite{PhysRevB.63.144425}. The second consequence of substituting the B-site cation involves subtle changes to the local oxygen environment surrounding each magnetic cation \cite{Subramanian198355}. These differences, although structurally small, can result in large differences to the single ion properties which determine the crystal electric field and, consequently, the anisotropy. 

In the case of some rare earth cations, the magnetic ground state is robust under the application of chemical pressure. For example, the spin ice state that was first observed in \pyro{Ho}{Ti}, persists at negative chemical pressure (\pyro{Ho}{Sn} \cite{0953-8984-12-40-103,PhysRevB.65.144421}) and positive chemical pressure (\pyro{Ho}{Ge} \cite{PhysRevLett.108.207206,PhysRevB.86.134431}). In other cases, chemical pressure can radically alter the magnetic ground state. For example, in the ytterbium based pyrochlores, \pyro{Yb}{B} (B = Sn, Ti, Ge), the interactions between ytterbium moments are very sensitive to changes in chemical pressure \cite{PhysRevB.89.064401}. Consequently, each of these pyrochlores have a markedly different magnetic ground state ranging from ferromagnetic Coulomb liquid in \pyro{Yb}{Sn} \cite{PhysRevB.87.134408} to antiferromagnetic order in \pyro{Yb}{Ge} \cite{PhysRevB.89.064401}. Different magnetic ground states are also observed in the terbium based pyrochlores, \pyro{Tb}{B} (B = Sn, Ti, Ge) \cite{Tb2Ge2O7,PhysRevLett.82.1012,PhysRevLett.94.246402}. %However, there is strong evidence to suggest that this difference is primarily due to crystal electric field effects \cite{PhysRevB.76.184436,PhysRevB.89.134410}.

Even more extreme differences in chemical pressure can be achieved through the substitution of lead onto the B-site of the pyrochlore lattice. Using high pressure synthesis techniques, the pyrochlore stability field was expanded in 1969 by A.~W.~Sleight to include the plumbates \cite{doi:10.1021/ic50078a060}. Pb$^{4+}$, with an ionic radius of 77.5~pm, is the largest B-site cation to be successfully incorporated into the pyrochlore structure \cite{doi:10.1021/ic50078a060}. However, the very large ionic radii of lead imposes limitations on possible A-site rare earth cations. Only the largest of the rare earths have been successfully prepared: \pyro{A}{Pb} (A = Pr, Nd, Gd). In fact, the only rare earth which can be prepared on both a titanate and plumbate lattice is gadolinium. Instead, the stannate, \pyro{A}{Sn}, and zirconate, \pyro{A}{Zr}, pyrochlores, are a more direct point of comparison. Each of the above mentioned rare earths can be combined with tin and zirconium onto the pyrochlore lattice. 

The praeseodymium pyrochlores, \pyro{Pr}{Sn} and \pyro{Pr}{Zr} play host to the so-called dynamic or quantum spin ice states \cite{PhysRevLett.101.227204,Pr2Zr2O7NatComm}. Much like the dipolar spin ice state, the Pr$^{3+}$ moments are locally arranged into a two-in/two-out configuration. However, the Pr$^{3+}$ moments are subject to stronger quantum fluctuations and remain dynamic down to the lowest measured temperatures, 200~mK \cite{PhysRevLett.101.227204}. This state is of significant experimental \cite{PhysRevLett.101.227204,Pr2Zr2O7NatComm,1742-6596-145-1-012031} and theoretical interest \cite{PhysRevB.86.104412,PhysRevLett.108.067204,PhysRevLett.105.047201,0034-4885-77-5-056501,PhysRevB.86.075154,PhysRevB.90.214430} as it may permit the propagation of monopole excitations. %The growth of large, high-quality single crystals of \pyro{Pr}{Zr} has facilitated . 
Neutron scattering measurements of single crystal \pyro{Pr}{Zr} have revealed a large inelastic signal, absent of pinch points, suggesting quantum fluctuations of magnetic monopoles \cite{Pr2Zr2O7NatComm}. 

There has been comparatively little study of the neodymium based pyrochlores. Long range ordering has been observed in \pyro{Nd}{Sn} and \pyro{Nd}{Zr} using heat capacity \cite{Blote} and magnetic susceptibility \cite{Stannates}. \pyro{Nd}{Sn} has a negative Weiss temperature, $\theta_{\text{CW}}$~=~$-0.31$~K \cite{Stannates}. However, until recently, the ground state of this material had not been determined. Recent magnetic neutron diffraction measurements have revealed the exact magnetic ground state in \pyro{Nd}{Sn} is an antiferromagnetic $q$~=~0 state \cite{Unpublished}. 

Gadolinium pyrochlores are unique in two respects. (i) Firstly, gadolinium has a spin only total angular momentum ($S$~=~7/2, $L$~=~0). As a result, the single ion anisotropy of Gd$^{3+}$ is small compared to other magnetic rare earth cations. (ii) Secondly, gadolinium pyrochlores have the largest size range for non-magnetic B-site cations, ranging from the smallest (Ge$^{4+}$ \cite{doi:10.1021/ic50066a038}) to the largest (Pb$^{4+}$ \cite{doi:10.1021/ic50078a060}). However, only the intermediate sized members, \pyro{Gd}{Ti} and \pyro{Gd}{Sn}, have been magnetically characterized until now \cite{PhysRevB.59.14489,BonvilleJPCM2003,PhysRevB.78.132410,PhysRevLett.89.067202}. Both \pyro{Gd}{Ti} and \pyro{Gd}{Sn} are candidates for the Palmer-Chalker ground state, a four sublattice N\'eel state \cite{PhysRevB.62.488}. \pyro{Gd}{Sn} does appear to be a good experimental realization of this state \cite{0953-8984-18-3-L02}. However, the exact ground state of \pyro{Gd}{Ti} has remained experimentally elusive \cite{PhysRevB.64.140407,0953-8984-16-28-L01,Gdunpublished}. 

With a diverse array of magnetic behavior, the frustrated pyrochlores have maintained a high level of interest over the past 20 years. An expanded knowledge of their phase diagrams will improve understanding of these unique ground states, and may allow the discovery of new ground states. Here, we report on the synthesis and characterization of the lead-based pyrochlores, \pyro{A}{Pb} (A~=~Pr, Nd, Gd). 

%********************************************************************************************
\section{Experimental Methods}
Polycrystalline samples of A$_2$Pb$_2$O$_7$ (A $=$ La, Nd, Pr, Gd) were synthesized from stoichiometric amounts of A$_2$O$_3$ and PbO$_2$ reacted in a gold capsule. The high pressure synthesis was carried out in a Walker-type multi-anvil press at 2 GPa and $900$$^{\circ}$C. Room temperature x-ray powder diffraction was performed with a D2-phaser diffractometer using Cu $K$-$\alpha$ radiation. Rietveld analyses of the x-ray diffraction patterns were performed using Fullprof \cite{RodriguezCarvajal1993}. Powder neutron diffraction measurements on a 90~mg sample of \pyro{Pr}{Pb} were made on the General Materials Diffractometer (GEM) at the ISIS neutron facility. Rietveld analysis of the neutron diffraction patterns were performed with GSAS.

The susceptibility and magnetization measurements were made using a Quantum Design MPMS magnetometer equipped with a $^3$He insert. The dc magnetic susceptibility measurements were made in fields of 1000~Oe between 0.5~K and 300~K. The ac magnetic susceptibility measurements were carried out with an ac driving field of 3.5~Oe, over a frequency range of 1~Hz to 1000~Hz. Specific heat measurements were carried out using a Quantum Design Physical Property Measurement System (PPMS) equipped with a $^3$He insert. The magnetic component of the specific heat was isolated by subtracting the specific heat of a non-magnetic lattice equivalent, \pyro{La}{Pb}.

%********************************************************************************************
\section{Crystal Structures}

\begin{figure}[htbp]
\linespread{1}
\par
\includegraphics[width=3.1in]{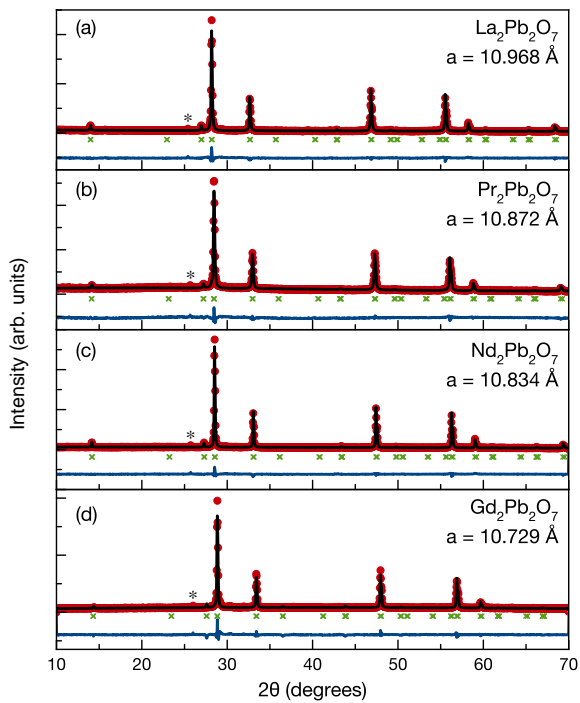}
\par
\caption{(Color online) Room temperature powder x-ray diffraction patterns and structure refinements of A$_2$Pb$_2$O$_7$, where A $=$ (a) La, (b) Nd, (c) Pr, and (d) Gd. Patterns for the observed (red marks), calculated (black lines), and difference (blue lines) are shown. The ticks indicate the positions of the Bragg reflections for the $Fd\bar{3}m$ space group.}
\label{Xrays}
\end{figure}

The crystal structures of the powder A$_2$Pb$_2$O$_7$ (A $=$ La, Nd, Pr, Gd) samples were investigated by room temperature powder x-ray diffraction. The crystal structure was determined by Rietveld refinement of the diffraction patterns using the space group $Fd\bar{3}m$ (Figure \ref{Xrays}). Lattice parameters for A$_2$Pb$_2$O$_7$ (A $=$ La, Nd, Pr, Gd), listed in Table~\ref{TableLattice}, were in agreement with prior investigations \cite{doi:10.1021/ic50078a060}. The lattice parameters for the plumbate pyrochlores follow a linear relation with A site ionic radii. The marked peak in each diffraction pattern at 2$\theta$~=~25.6$^{\circ}$ corresponds to the (222) pyrochlore reflection from the Cu $K$-$\beta$ wavelength.

\begin{table}[tbp]
\caption{Ionic radii for A = Pr, Nd, Gd and B = Ti, Sn, Pb, and the resultant pyrochlore lattice parameter. The lattice parameters for the stannate pyrochlores are taken from \cite{KennedyPyro}.}
\begin{tabular}{>{\centering\arraybackslash}p{1.9cm}|>{\centering\arraybackslash}p{1.9cm}>{\centering\arraybackslash}p{1.9cm}>{\centering\arraybackslash}p{1.9cm}}
\toprule
 Cation & Pr (0.99 \AA) & Nd (0.98 \AA) & Gd (0.94 \AA) \\
\colrule
Ti (0.61 \AA) & - & - & 10.18 \AA \cite{BonvilleJPCM2003} \\
Sn (0.69 \AA) & 10.600 \AA & 10.567 \AA & 10.454 \AA \\
Pb (0.78 \AA) & 10.872 \AA & 10.834 \AA & 10.729 \AA \\
\botrule
\end{tabular}
\label{TableLattice}
\end{table}

\begin{figure}[htbp]
\linespread{1}
\par
\includegraphics[width=3.1in]{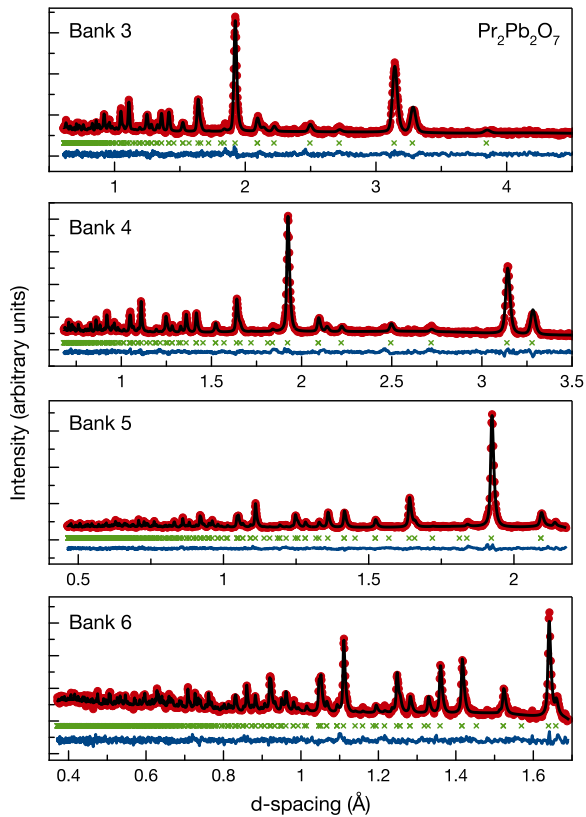}
\par
\caption{(Color online) Rietveld refinement of the room temperature TOF diffraction pattern of \pyro{Pr}{Pb} measured on the GEM diffractometer. The observed (red marks), calculated (black lines) and difference (blue lines) patterns are shown. The ticks indicate the positions of the Bragg reflections for the $Fd\bar{3}m$ space group. The weighted profile R-factor for the fit is $R_{wp}~=~4.14\%$.}
\label{Neutrons}
\end{figure}

Lead is the largest known possible B-site cation for the pyrochlore lattice. As a result, the A-site to B-site ratio, $A^{3+}/B^{4+}$, in the lead pyrochlores are amongst the smallest. Consequently, much like zirconium-based pyrochlores, non-negligible site-mixing between the A and B sites is expected to be likely \cite{doi:10.1021/ic301677b}. To determine the exact level of site-mixing in the lead pyrochlores, we have performed room temperature neutron diffraction on \pyro{Pr}{Pb} (Figure~\ref{Neutrons}). Rietveld analysis revealed the presence of mixing between Pr$^{3+}$ and Pb$^{4+}$ on the order of 8\%. %In the zirconium pyrochlores, there is also an increasing tendency towards the defect fluorite structure with decreasing rare earth cation size. It is likely that this is also a factor in the lead-pyrochlores. 
This is also evidenced by the reduced size of the (111) superlattice peak at 2$\theta$~=~14$^{\circ}$ with decreasing rare earth size (Figure~\ref{Xrays}). The pyrochlore superlattice peaks arise due to A and B-site ordering, and vanish in the case of random site occupancy. The detailed results of the structural refinement of \pyro{Pr}{Pb} are presented in Table~\ref{NeutronRefine}.

\begin{table}[tbp]
\caption{Selected parameters for the refinement of the neutron diffraction pattern of \pyro{Pr}{Pb} }
\begin{tabular}{>{\centering\arraybackslash}p{1.1cm}|>{\centering\arraybackslash}p{0.9cm}>{\centering\arraybackslash}p{1.8cm}>{\centering\arraybackslash}p{2.4cm}>{\centering\arraybackslash}p{1.6cm}}
\toprule
Atom & Site & (x,y,z) & Occupancy & U$_{iso}$ (\AA$^2$) \\
\colrule
Pr/Pb & $16d$ & $(\frac{1}{2},\frac{1}{2},\frac{1}{2})$ & 0.924(7)/0.076(7) & 0.0071(9) \\
Pb/Pr & $16c$ & (0,0,0) & 0.924(7)/0.076(7) & 0.0026(4) \\
O1 & $8b$ & $(\frac{3}{8},\frac{3}{8},\frac{3}{8})$ & 0.91(1) & 0.006(1) \\
O2 & $48f$ & $(0.342,\frac{1}{8},\frac{1}{8})$ & 1.0 & 0.0085(4) \\
\botrule
\end{tabular}
\label{NeutronRefine}
\end{table}

\pyro{Gd}{Pb} has an A-site to B-site ratio of 1.36, the smallest of any pyrochlore \cite{GGG}. From inspection of the x-ray diffraction pattern, it is apparent that the (111) superlattice peak at 2$\theta$~=~14$^{\circ}$ is nearly extinct in \pyro{Gd}{Pb} (Figure~\ref{Xrays}(d)). However, neutron diffraction of \pyro{Gd}{Pb} is made difficult by the high neutron absorption cross section of gadolinium. Instead, we used x-ray refinements to investigate the degree of site-mixing in \pyro{Gd}{Pb}. Refinements in which the A-site and B-site occupancies are allowed to freely vary give 61\% order of the A and B sites (39\% site mixing), close to the random occupancies (50\%) expected for the defect-fluorite structure. For comparison, x-ray refinements of the other three lead pyrochlores we investigated, \pyro{A}{Pb} (A~=~La, Nd, Pr) each gave 96\% order (4\% site mixing), which agrees well with the neutron refinements.

%********************************************************************************************
\section{Magnetic Properties: P\lowercase{r}$_2$P\lowercase{b}$_2$O$_7$}

\begin{figure}[tbp]
\linespread{1}
\par
\includegraphics[width=3.1in]{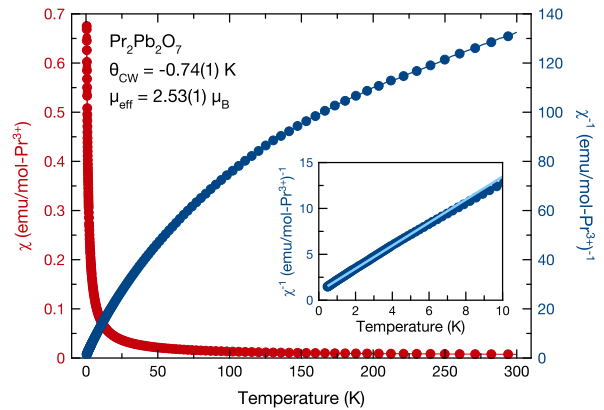}
\par
\caption{(Color online) The temperature dependence of the magnetic susceptibility (red) and inverse susceptibility (blue) of Pr$_2$Pb$_2$O$_7$ measured with a field of 1000~Oe. Inset: Low temperature inverse susceptibility with a Curie-Weiss law fit between 0.5~K and 5~K.}
\label{PrPbO_Susc}
\end{figure}

Figure \ref{PrPbO_Susc} shows the magnetic susceptibility of Pr$_2$Pb$_2$O$_7$ measured in a field of 1000~Oe. The susceptibility of Pr$_2$Pb$_2$O$_7$ is free of anomalies down to 0.5~K. The inverse susceptibility has a large negative curvature with increasing temperature, likely due to a large splitting of the crystal electric field levels. Consequently, the region over which the Curie-Weiss law can be applied is limited to low temperatures, between 0.5~K and 5~K (Figure~\ref{PrPbO_Susc} inset). The effect of low lying crystal fields is to limit the quantitative validity of the Curie-Weiss analysis; however, we proceed with the intention of drawing comparisons with other praeseodymium pyrochlores. The fit gives a Weiss temperature of $\theta_{\text{CW}}$~=~$-0.74(1)$~K, indicating net antiferromagnetic interactions. The Weiss temperatures in \pyro{Pr}{Sn} and \pyro{Pr}{Zr} are 0.32~K \cite{Stannates} and $-1.41$~K \cite{Pr2Zr2O7NatComm}, respectively. However, the temperature regions which exhibit Curie-Weiss behavior differs significantly between these three compounds. As a result, it is difficult to draw comparisons on the basis of the Weiss temperature. The strength of the nearest neighbor dipolar interactions, which can be approximated in pyrochlores by:
\begin{equation}
D_{\text{nn}}~=~\frac{5\mu_0}{3\cdot4\pi}\frac{\mu^2}{r_{nn}^{3}}
\end{equation}  
when $\mu$ is estimated from the low temperature susceptibility, is only 0.11~K in \pyro{Pr}{Pb}. Thus, the magnetism in \pyro{Pr}{Pb} is dominated by exchange interactions. 

The higher temperature magnetic susceptibility of \pyro{Pr}{Pb} is dominated by single ion physics, due to a large crystal electric field splitting. The 9-fold degenerate state of Pr$^{3+}$ is split into three non-Kramers doublets and three singlets \cite{Stannates}. From the Curie-Weiss fit between 0.5 and 5 K, we calculate an effective magnetic moment of 2.53(1) $\mu_{\text{B}}$, a value not significantly different from that found in \pyro{Pr}{Sn} and \pyro{Pr}{Zr} \cite{Stannates,Pr2Zr2O7NatComm}. However, this moment falls short of the 3.58 $\mu_{\text{B}}$ free-ion moment for a pure $\ket{J=\pm4}$ state, suggesting a mixing of $\ket{J=\pm1}$ and $\ket{J=\mp2}$ terms \cite{Stannates}. Thus, \pyro{Pr}{Pb} has a doublet ground state with Ising moments pointing along the local $<$111$>$ trigonal axes. 
 
\begin{figure}[tbp]
\linespread{1}
\par
\includegraphics[width=3.1in]{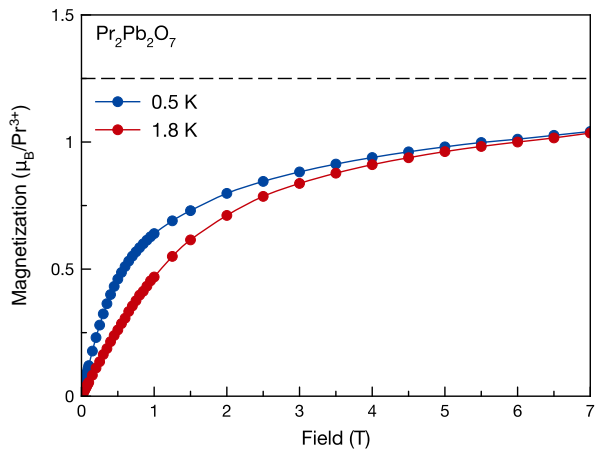}
\par
\caption{(Color online) The magnetization of Pr$_2$Pb$_2$O$_7$ as a function of applied field between 0~T and 7~T at 0.5~K and 1.8~K. The dashed line marks half the effective moment value, 1.25~$\mu_{\text{B}}$/Pr.}
\label{PrPbO_Mag}
\end{figure}

The magnetization of Pr$_2$Pb$_2$O$_7$, shown in Figure \ref{PrPbO_Mag}, does not saturate in a field of 7~T at 0.5~K. This behavior deviates strongly from the Brillouin function for $J$~=~4, 
\begin{equation}
B_J(x) = \frac{2J+1}{2J} \coth{\left( \frac{2J+1}{2J}x \right)} - \frac{1}{2J} \coth{\left( \frac{x}{2J} \right)}
\end{equation}
\begin{equation}
\text{where } \text{ } \text{ } \text{ } \text{ } x = \frac{g_J\mu_B J B}{k_B T}
\end{equation}
due to significant interactions between the Pr$^{3+}$ moments at these temperatures and deviations from isotropic crystal field levels. The magnetization of \pyro{Pr}{Pb} appears to be approaching a saturation value of $\sim$1.25~$\mu_{\text{B}}$~=~$\frac{\mu_{\text{eff}}}{2}$. A saturation value of half the effective moment in powder samples is a hallmark of $<$111$>$ Ising anisotropy, as observed in other praseodymium pyrochlores \cite{Matsuhira2004E981,KimuraJKor2013}. In general, the magnetization of \pyro{Pr}{Pb} closely resembles that of \pyro{Pr}{Zr}, which also fails to saturate by 7~T \cite{KimuraJKor2013}. %The magnetization of \pyro{Pr}{Sn} differs significantly from its group members. Conversely, in \pyro{Pr}{Sn} the magnetization sharply increases to above 1~$\mu_{\text{B}}$/Pr by 1~T and saturates to half the effective moment, 1.26~$\mu_{\text{B}}$~=~$\frac{\mu_{\text{eff}}}{2}$ by 7~T \cite{Matsuhira2004E981}.

\begin{figure}[tbp]
\linespread{1}
\par
\includegraphics[width=3.1in]{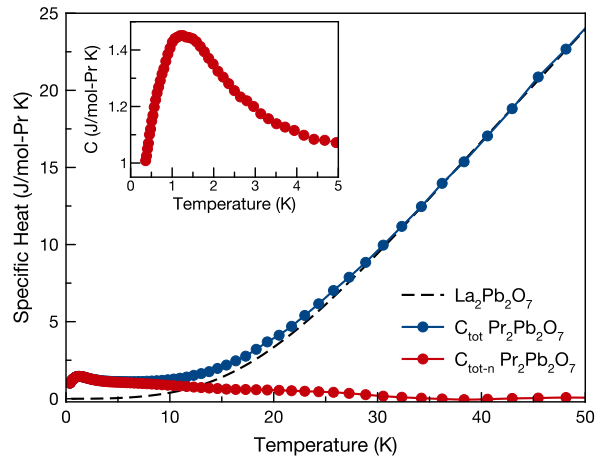}
\par
\caption{(Color online) The specific heat of \pyro{Pr}{Pb} and non-magnetic \pyro{La}{Pb}. Inset: low temperature lattice subtracted specific heat.}
\label{PrPbO_Cp}
\end{figure}

The heat capacity of \pyro{Pr}{Pb} contains a broad anomaly centered at 1.2~K (Figure~\ref{PrPbO_Cp}). This feature does not resemble the sharp anomalies which typically signify long-range ordering, nor does it fit to a Schottky model. Instead, the peak in the heat capacity of \pyro{Pr}{Pb} bears a resemblance to those observed in spin ice, attributed to a condensation of thermally activated monopole pairs \cite{Pr2Zr2O7NatComm}. Much like \pyro{Pr}{Zr}, the peak is broader than those observed in the dipolar spin ices \cite{Ramirez1999}. This broadening is attributed to monopole quantum dynamics in \pyro{Pr}{Zr}, and is likely also at play in \pyro{Pr}{Pb}. 

The lattice contribution to the specific heat was subtracted by using the specific heat of non-magnetic \pyro{La}{Pb}. Calculation of the entropy for the lattice subtracted specific heat, $\Delta S~=~\int \frac{C_{p}}{T}dT$, gives a value which exceeds the maximum value for an Ising system, $Rln(2)$. However, there are other known contributions to the specific heat of \pyro{Pr}{Pb}. The hyperfine splitting of the nuclear levels of Pr$^{3+}$ produces a nuclear Schottky anomaly with significant intensity below 1~K. %Without capturing the upturn of the nuclear Schottky, subtraction is not possible. 
Furthermore, the crystal field scheme of \pyro{Pr}{Pb} has not been measured, but there is likely a low-lying crystal field contributing to the specific heat, as evidenced from changes in the curvature close to 10~K. However, to subtract this contribution would require a more precise knowledge of the low-lying crystal field levels. Since these contributions to the specific heat of \pyro{Pr}{Pb} cannot be accurately determined, a realistic calculation of the magnetic entropy is not possible. It is interesting to note that the lowest crystal fields in \pyro{Pr}{Sn} and \pyro{Pr}{Zr} occur at energies an order of magnitude higher \cite{Pr2Zr2O7NatComm,PhysRevB.88.104421}. Thus, \pyro{Pr}{Pb} represents an interesting opportunity to investigate the impact of a low-lying crystal field on a dynamic spin ice state.

%********************************************************************************************
\section{Magnetic Properties: N\lowercase{d}$_2$P\lowercase{b}$_2$O$_7$}

\begin{figure}[tbp]
\linespread{1}
\par
\includegraphics[width=3.1in]{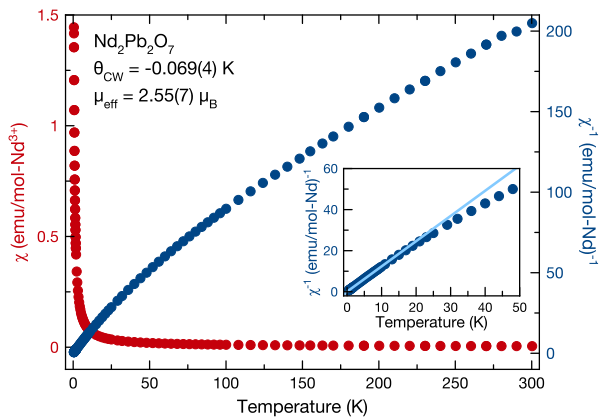}
\par
\caption{(Color online) The magnetic susceptibility of \pyro{Nd}{Pb} measured with an applied field of 1000~Oe. The left axis gives the direct susceptibility (red) and the right axis gives the inverse susceptibility (blue). The Curie-Weiss fit between 0.5~K and 15~K is given by the blue line.}
\label{NdPbO_Susc}
\end{figure}

The magnetic susceptibility of Nd$_2$Pb$_2$O$_7$, shown in Figure \ref{NdPbO_Susc}, contains no anomalies down to 0.5~K. Between 0.5~K and 15~K, the inverse susceptibility is well-fit by the Curie-Weiss law. Before proceeding, we should note that much like  Pr$_2$Pb$_2$O$_7$, the crystal electric field is likely impacting the analysis. The Weiss temperature calculated from this fit is $\theta_{\text{CW}}$ = $-$0.069(4)~K, indicating weak net antiferromagnetic interactions. A Curie-Weiss fit over the same temperature range in \pyro{Nd}{Sn} gives a Weiss temperature of $-0.31$~K \cite{Stannates}. %\pyro{Nd}{Zr} has a reported Curie-Weiss temperature of 0.06(5)~K \cite{Blote}.
Thus, the change from the stannate to the plumbate results in a very close competition of antiferromagnetic and ferromagnetic interactions. It is worth noting that the calculated dipolar interactions in \pyro{Nd}{Pb} and \pyro{Nd}{Sn} are 0.11~K and 0.12~K, respectively. Thus, the changing strength of the dipolar interaction alone cannot account for the change in the Weiss temperature. The decreased Weiss temperature in \pyro{Nd}{Pb} is thus indicative of decreased exchange interactions. 

\begin{figure}[tbp]
\linespread{1}
\par
\includegraphics[width=3.1in]{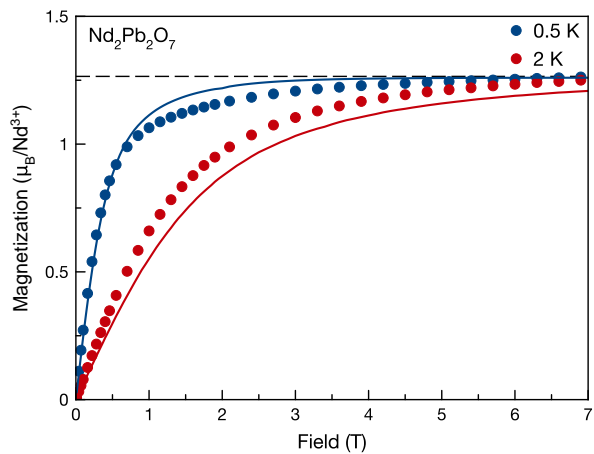}
\par
\caption{(Color online) The magnetization of Nd$_2$Pb$_2$O$_7$ as a function of applied field between 0~T and 7~T at 0.5~K and 2~K. The solid blue and red curves give the Brillouin function for 0.5~K and 2~K respectively.}
\label{NdPbO_Mag}
\end{figure}

The calculated effective magnetic moment for \pyro{Nd}{Pb}, from the Curie-Weiss fit, is 2.55(7)~$\mu_{\text{B}}$. This value falls short of the free-ion value of 3.62~$\mu_{\text{B}}$ for $\ket{J=\pm9/2}$, indicating the mixing of $\ket{J=\pm3/2}$ terms. Above 20~K, the susceptibility deviates from Curie-Weiss behavior due to low lying crystal field levels. The 10-fold degenerate ground state of Nd$^{3+}$ ($J=9/2$) is split into five symmetry protected Kramers doublets. Figure \ref{NdPbO_Mag} shows the magnetization of \pyro{Nd}{Pb} at 0.5~K and 2~K. At both temperatures, the magnetization is similar to that of a Brillouin function for J~=~9/2. The magnetization saturates at 1.26~$\mu_{\text{B}}$/Nd$^{3+}$, half the value of the effective moment. Again, this is a consequence of local $<$111$>$ Ising anisotropy and powder averaging. 

\begin{figure}[tbp]
\linespread{1}
\par
\includegraphics[width=3.1in]{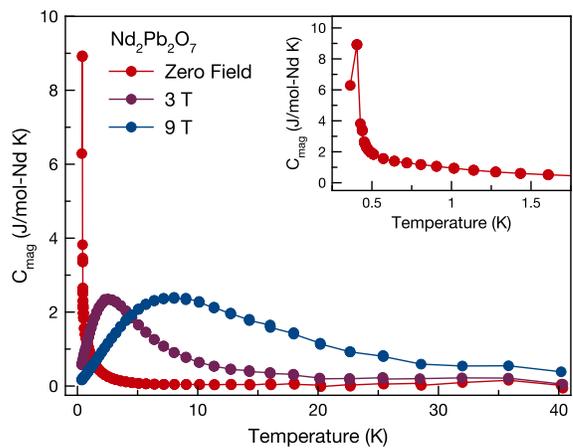}
\par
\caption{(Color online) Magnetic specific heat of \pyro{Nd}{Pb} measured in zero field, 3~T, and 9~T between 0.4~K and 40~K. Inset: Scaled in view of the low temperature zero field specific heat of \pyro{Nd}{Pb}.}
\label{NdPbO_Cp}
\end{figure}

\begin{figure}[tbp]
\linespread{1}
\par
\includegraphics[width=3.1in]{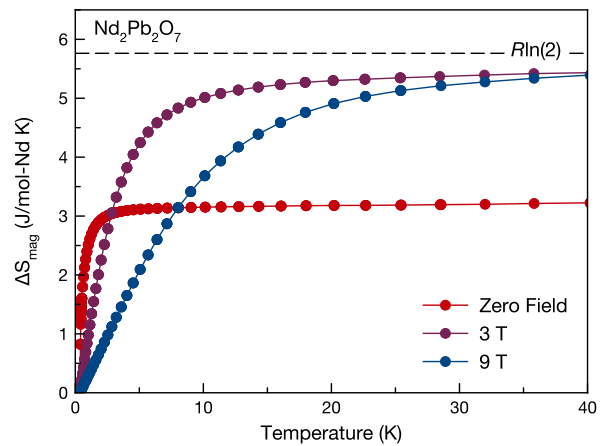}
\par
\caption{(Color online) Calculated magnetic entropy of \pyro{Nd}{Pb} measured in zero field, 3~T, and 9~T between 0.4~K and 40~K. The dashed line indicates the expected value, $R\ln{(2)}$, for an Ising system.}
\label{NdPbO_Entropy}
\end{figure}

The magnetic specific heat of \pyro{Nd}{Pb} is shown in Figure \ref{NdPbO_Cp}. In zero field, the magnetic component of the specific heat increases below 3~K. Below 0.5~K, the specific heat begins to strongly increase, with its maximum occurring at 0.41~K. However, only a single data point could be collected below the peak's maximum. This peak, which is likely due to long-range magnetic ordering, shifts upwards in temperature and broadens with an externally applied field. A field of 3~T shifts the maximum to 2.5~K, while a field of 9~T further increases it to 8~K. The magnetic origin of this peak is most easily verified by integrating out the entropy, which should give $R\cdot\ln{2}$ for an Ising system. However, this cannot be done with our zero-field measurement, in which the decreasing side of the anomaly is outside our measurement range (Figure~\ref{NdPbO_Cp}). Integration of $C_p/T$ between 0.4~K and 40~K for the 3~T and 9~T data sets yields 0.95$\cdot$$R\ln{2}$ and 0.96$\cdot$$R\ln{2}$, respectively (Figure~\ref{NdPbO_Entropy}). %Furthermore, the region above the peak fits well to a logarithmic divergence function (Figure \ref{NdPbO_Cp}) inset):
%\begin{equation}
%A-B\cdot\ln{|1-T/T_c|}
%\end{equation}
%A logarithmic divergence of specific heat with temperature is consistent with a lambda-type anomaly, further evidence for a magnetic ordering transition. Thus, we can conclude that \pyro{Nd}{Pb} is undergoing long-range magnetic ordering at 0.405~K. 

Long-range order was also found in \pyro{Nd}{Sn}, which has both a sharp peak in its heat capacity \cite{Blote} and a cusp in its susceptibility \cite{Stannates} at 0.9~K. %The magnetism of \pyro{Nd}{Zr} has not been as extensively characterized. The specific heat of \pyro{Nd}{Zr} contains a peak centered at 0.37~K, which is broadened due to the presence of chemical inhomogeneities \cite{Blote}. 
It is interesting to note that the ordering temperature in \pyro{Nd}{Pb} is reduced from \pyro{Nd}{Sn} by a factor of two. Recent neutron scattering measurements of \pyro{Nd}{Sn} have confirmed that the ordered state is the $k$=0 antiferromagnetic all-in/all-out state \cite{Unpublished}. Neutron scattering will also be required to determine the consequence of net ferromagnetic interactions on the magnetic ground state of \pyro{Nd}{Pb}.

%********************************************************************************************
\section{Magnetic Properties: G\lowercase{d}$_2$P\lowercase{b}$_2$O$_7$}

\begin{figure}[tbp]
\linespread{1}
\par
\includegraphics[width=3.1in]{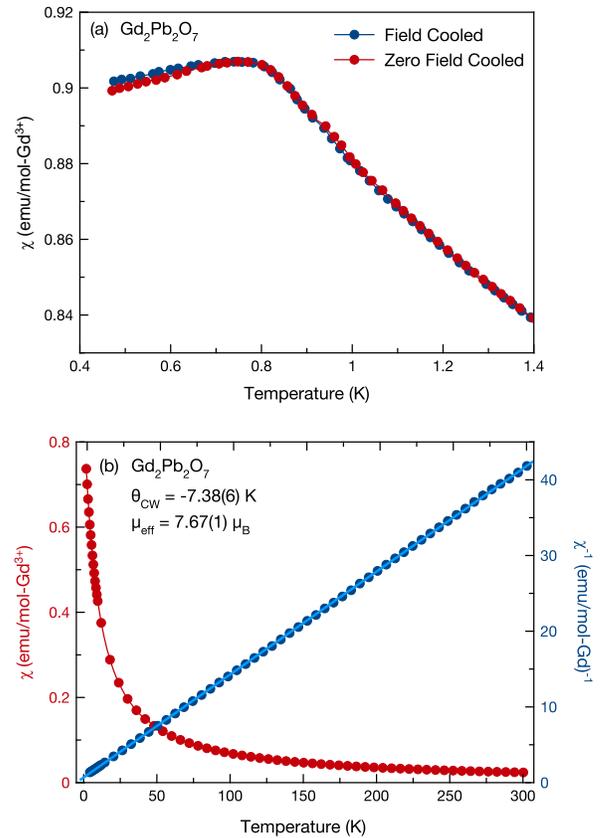}
\par
\caption{(Color online) (a) The low temperature magnetic susceptibility of Gd$_2$Pb$_2$O$_7$ measured with an applied field of 1000 Oe. Below the magnetic ordering transition at 0.8~K there is a splitting between the zero field cooled and field cooled susceptibility. (b) The high temperature magnetic susceptibility of Gd$_2$Pb$_2$O$_7$. The left axis gives the direct susceptibility (red) and the right axis gives the inverse susceptibility (blue). The Curie-Weiss law fit is given by the blue line.}
\label{GdPbO_Susc}
\end{figure}

Figure \ref{GdPbO_Susc}(a) shows the low temperature magnetic susceptibility of Gd$_2$Pb$_2$O$_7$ measured in 1000 Oe. A magnetic ordering transition is signaled by a cusp in the susceptibility at 0.8~K. There is a slight splitting of the field cooled and zero field cooled susceptibility below the transition at 0.8~K indicating a small amount of irreversibility. The susceptibilities of the related pyrochlores, Gd$_2$Ti$_2$O$_7$ and Gd$_2$Sn$_2$O$_7$, both contain magnetic ordering anomalies close to 1.0~K \cite{BonvilleJPCM2003}. They too both exhibit a degree of irreversibility below these transitions. The maximum FC/ZFC splitting measured in Gd$_2$Pb$_2$O$_7$ is 0.03~emu/mol-Gd$^{3+}$, very similar to the magnitude of the splitting in Gd$_2$Ti$_2$O$_7$ \cite{BonvilleJPCM2003}. Conversely, in Gd$_2$Sn$_2$O$_7$, the FC/ZFC splitting is an order of magnitude larger \cite{BonvilleJPCM2003}.  %However, unlike Gd$_2$Ti$_2$O$_7$, there is no evidence of a second smaller anomaly in the susceptibility of Gd$_2$Pb$_2$O$_7$.

The high temperature magnetic susceptibility of Gd$_2$Pb$_2$O$_7$ up to 300~K is shown in Figure \ref{GdPbO_Susc}(b). Between 10~K and 300~K the inverse susceptibility is strikingly linear, fitting well to Curie-Weiss behavior. The Weiss temperature is $-7.38(6)$~K, indicating dominant antiferromagnetic interactions. The Weiss temperatures for Gd$_2$Ti$_2$O$_7$ and Gd$_2$Sn$_2$O$_7$ are $-$9.9(1)~K and $-$8.6(1)~K, respectively \cite{BonvilleJPCM2003}. As anticipated, the substitution of the larger Pb$^{4+}$ cation appears to have weakened the antiferromagnetic exchange in Gd$_2$Pb$_2$O$_7$.

The Curie-Weiss fit of Gd$_2$Pb$_2$O$_7$ can also be used to extract an effective moment, yielding 7.67(1)~$\mu_{\text{B}}$ (Figure \ref{GdPbO_Susc}(b)). This calculated moment is close to the expected 7.94~$\mu_{\text{B}}$ free-ion moment. In Gd$^{3+}$, the isotropic spin, $S~=~7/2$, is the only contribution to the total angular momentum, $J$. Consequently, the 8-fold degeneracy of the ground state is only marginally lifted by the crystal electric field, on the order of a fraction of a Kelvin \cite{PhysRevB.59.14489}. The high degree of linearity in the inverse susceptibility between 10~K to 300~K suggests there are no well-separated crystal field levels in Gd$_2$Pb$_2$O$_7$, supporting the picture described above. 

\begin{figure}[tbp]
\linespread{1}
\par
\includegraphics[width=3.1in]{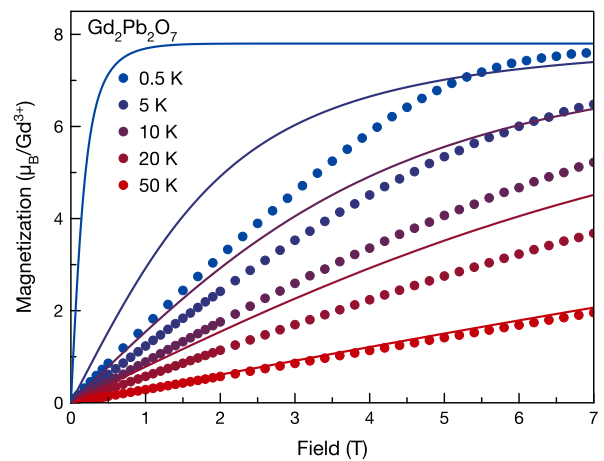}
\par
\caption{(Color online) The magnetization of Gd$_{2}$Pb$_{2}$O$_{7}$ as a function of applied field between 0~T and 7~T at temperatures between 0.5~K and 50~K. The solid curves show the Brillouin function for $J = 7/2$ at each corresponding temperature.}
\label{GdPbO_Mag}
\end{figure}

The magnetization of Gd$_2$Pb$_2$O$_7$ is shown in Figure \ref{GdPbO_Mag} at temperatures ranging from 0.5~K to 50~K. At 10~K and higher, the magnetization fits well to the Brillouin function for $J~=~7/2$, which describes a system of paramagnetic, non-interacting moments. At 0.5~K, where the Brillouin function fails completely, the magnetization approaches a saturation value on the order of the free-ion Gd$^{3+}$ moment by 7~T. 

\begin{figure}[tbp]
\linespread{1}
\par
\includegraphics[width=3.1in]{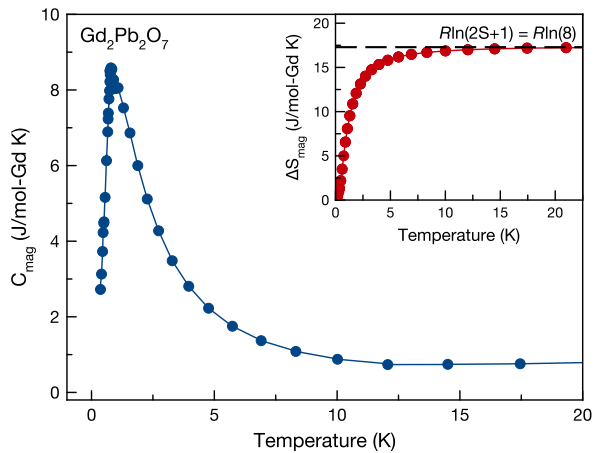}
\par
\caption{(Color online) Magnetic specific heat of \pyro{Gd}{Pb} measured in zero field between 0.4~K and 20~K. Inset: Integrated entropy of \pyro{Gd}{Pb}, giving the expected value, $R\ln{8}$, for Heisenberg-like spins.}
\label{GdPbO_Cp}
\end{figure}

The magnetic specific heat of \pyro{Gd}{Pb} is shown in Figure \ref{GdPbO_Cp}. The magnetic specific heat increases below 10~K and peaks at 0.81~K. Extrapolating $C_{\text{mag}}/T$ to 0~J/mol-K$^2$, gives an integrated entropy of 17.1~J/mol-K (Figure~\ref{GdPbO_Cp} inset). This value corresponds to 0.99$\cdot$$R\ln{8}$~=~0.99$\cdot$$R\ln{(2S+1)}$. This is the expected value for Gd$^{3+}$ ($S$~=~$7/2$), which, with very small single ion anisotropy, has Heisenberg-like spins. Anomalies related to magnetic ordering are also observed in \pyro{Gd}{Ti} and \pyro{Gd}{Sn} at approximately 1~K. The maximum amplitude of the anomaly in \pyro{Gd}{Pb} is 8.7~J/mol-K. This is similar in magnitude to that of \pyro{Gd}{Ti}, whereas the anomaly in \pyro{Gd}{Sn} has a maximum amplitude of 120~J/mol-K \cite{BonvilleJPCM2003}. However, in \pyro{Gd}{Ti}, there is a second anomaly at 0.75~K and the ordering in this compound occurs in two distinct stages \cite{PhysRevLett.89.067202,BonvilleJPCM2003}. Thus, the heat capacity anomaly of \pyro{Gd}{Pb} does not strongly resemble either of its group members. 

\begin{figure}[tbp]
\linespread{1}
\par
\includegraphics[width=3.1in]{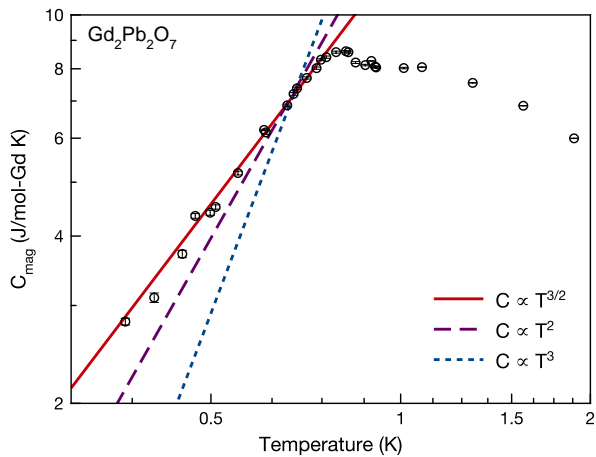}
\par
\caption{(Color online) Power law fits to the low temperature specific heat of \pyro{Gd}{Pb}. The least squares refinement gives a $C_{\text{mag}} \propto T^{3/2}$ dependence (red curve).}
\label{GdPbO_PowerLaw}
\end{figure}

The temperature dependence of the magnetic specific heat of \pyro{Gd}{Pb}, below the ordering temperature, reveals further differences from the other gadolinium pyrochlores (Figure~\ref{GdPbO_PowerLaw}). Conventional ungapped antiferromagnetic spin wave excitations give rise to a $C_{\text{mag}} \propto T^3$. However, the specific heat below the ordering transition in both \pyro{Gd}{Ti} and \pyro{Gd}{Sn} is better parameterized by $C_{\text{mag}} \propto T^2$ \cite{PhysRevLett.95.047203,BonvilleJPCM2003}. In \pyro{Gd}{Sn}, the specific heat is found to decrease exponentially below 350~mK \cite{PhysRevLett.99.097201}. This behavior in \pyro{Gd}{Sn} is attributed to conventional antiferromagnetic gapped magnons and can be theoretically reconciled with linear spin-wave theory \cite{PhysRevLett.99.097201}. However, the $C_{\text{mag}} \propto T^2$ dependence in \pyro{Gd}{Ti}, which is ungapped down to 100~mK, remains a mystery \cite{PhysRevLett.95.047203}. 

A least squares fit to the heat capacity of \pyro{Gd}{Pb} between 0.35~K and 0.75~K to $C_{\text{mag}} \propto T^{\eta}$ gives $\eta~=~1.49(3)$. Thus, the specific heat of \pyro{Gd}{Pb} below 0.8~K is well described by a $C_{\text{mag}} \propto T^{3/2}$ power law, as shown in Figure~\ref{GdPbO_PowerLaw}. In an ordered ferromagnet, spin wave excitations make a $C_{mag}~\propto~T^{3/2}$ contribution to the specific heat. However, a ferromagnetic ground state cannot be reconciled with the overwhelmingly negative (antiferromagnetic) Curie-Weiss temperature in Gd$_2$Pb$_2$O$_7$. Determination of the magnetic ground state of Gd$_2$Pb$_2$O$_7$ will require neutron diffraction, which is made challenging by the high neutron absorption cross section of gadolinium. While the exact ground state remains unknown at present, it is clear that it is markedly different from those of \pyro{Gd}{Ti} and \pyro{Gd}{Sn}. In the absence of neutron diffraction or muon spin relaxation microscopic magnetic measurements we cannot rule out spin glass freezing. However, the sharpness of the transition seen in our bulk dc suscepbitility and specific heat measurements, along with a lack of frequency dependence in ac susceptibility (not shown) make spin glass ordering highly unlikely in \pyro{Gd}{Pb}. This is particularily noteworthy given the significant degree of site mixing.

\section{Conclusions}

\begin{table}[tbp]
\caption{Comparison of lattice parameters and selected magnetic parameters in the A$_2$Pb$_2$O$_7$ (A = La, Nd, Pr and Gd pyrochlores).}
\begin{tabular}{lcccccccc}
\toprule
 && $a$ (\AA) && $\theta_{\text{CW}}$ (K) && $\mu_{\text{eff}}$ ($\mu_{\text{B}}$) && Anisotropy\\
\colrule
La$_2$Pb$_2$O$_7$ && 10.9682(4) && $-$ && $-$ && $-$ \\
Pr$_2$Pb$_2$O$_7$ && 10.8721(9) && $-0.74(1)$ && 2.53(1) && Ising \\
Nd$_2$Pb$_2$O$_7$ && 10.8336(4) && $0.15(1)$ && 2.47(1) && Ising \\
Gd$_2$Pb$_2$O$_7$ && 10.7292(8) && $-7.38(6)$ && 7.67(1) && Heisenberg \\
\botrule
\end{tabular}
\label{PbProperties}
\end{table}

We have reported on the crystallographic and magnetic properties of the lead based pyrochlores \pyro{A}{Pb} (A = La, Pr, Nd, Gd) (Table~\ref{PbProperties}). We have explored a new region of phase space in a family of materials known to have a rich collection of ground states. Lead is the largest B-site cation which can occupy the pyrochlore lattice. Consequently, lead-substitution on the pyrochlore lattice results in an effective weakening of the exchange interaction. As a result, magnetic ordering is suppressed to lower temperatures in each case. \pyro{Pr}{Pb} has strong Ising $<$111$>$ anisotropy and an absence of long-range magnetic order down to 500~mK. \pyro{Pr}{Pb} has a spin ice-like anomaly in its heat capacity centered at 1.2~K. Thus, \pyro{Pr}{Pb} is a candidate for a quantum spin ice state, despite weakened exchange and mild disorder. \pyro{Nd}{Pb} also has strong $<$111$>$ Ising anisotropy. However, \pyro{Nd}{Pb} transitions to long-range magnetic order at 0.41~K. The small, negative Weiss temperature of \pyro{Nd}{Pb} indicates it may lie on the border between ferromagnetism and antiferromagnetism. \pyro{Gd}{Pb} is a good approximation to Heisenberg spins on the pyrochlore lattice. However, due to the very similar ionic radii of gadolinium and lead, \pyro{Gd}{Pb} has a significant level of site mixing. Despite this structural disorder, magnetic ordering in \pyro{Gd}{Pb} at 0.81~K is signaled by a cusp in the susceptibility and a lambda-like specific heat anomaly. Unlike \pyro{Gd}{Ti} the ordering in \pyro{Gd}{Pb} occurs in a single stage. Furthermore, a $T^{3/2}$ dependence is observed in the low temperature specific heat, a value which is typically associated with the spin wave contribution in a conventional ferromagnet. Thus, it is unlikely that \pyro{Gd}{Pb} is adopting the antiferromagnetic Palmer-Chalker ground state. To that end, neutron diffraction of \pyro{Gd}{Pb} is recommended. In general, these lead-based pyrochlores provide insight into the effects of weaker exchange on highly frustrated ground states. 

\begin{acknowledgments}
We thank Dr. W. Kockelmann (ISIS) for assistance with diffraction measurements and Michel Gingras for a careful reading of our manuscript. This work was supported by the Natural Sciences and Engineering Research Council of Canada, the Canada Research Chairs program, and the Canadian Foundation for Innovation. We acknowledge support from the Engineering and Physical Sciences Research Council, the Science and Technology Facilities Council, and the Royal Society. A.M.H. is grateful for support from the Vanier Canada Graduate Scholarship program.
\end{acknowledgments}

\bibliography{PbPyrochloreRef}

\end{document}